\documentclass[a4paper]{jpconf}
\usepackage{graphicx}
\usepackage{amsmath,amsthm,amssymb,latexsym}
\begin{document}
\title{On the stability of homogeneous black strings in AdS}

\author{Carla Henr\'iquez-B\'aez}

\address{Departamento de F\'isica, Universidad de Concepci\'on, Casilla 160-C, Concepci\'on, Chile}

\ead{carlalhenriquez@udec.cl}

\begin{abstract}
In this paper we extend the analysis of the stability of an homogeneous black string in the presence of a negative cosmological constant with minimally coupled scalar fields. We recall the linear stability of this solutions under generic perturbations on the metric and of the scalar fields. Then, we extend the study of the stability by presenting the existence of a non-generic perturbation which may lead to an unstable behavior. The later mode is fine-tuned since it requires the scalar field degree of freedom to be absent through the whole evolution of the system.

\end{abstract}

\section{Introduction}

In higher dimensions, black holes, or black objects, exhibit different properties than in the four-dimensional scenario. For example, uniqueness theorems change, event horizon topology constraints no longer hold, among others \cite{Horowitz}, \cite{Obers}. In particular, it was shown that black string solutions are unstable under long wavelenght perturbations \cite{GL93}, phenomenon which is known as Gregory-Laflamme instability (GL). The evolution of this instability will lead to a pinch-off of the horizon making visible the central singularity \cite{Lenher-Pretorius}. This type of instability is present in a wide variety of black objects \cite{Inst.bs}, such as rotating black rings, rotating black holes and also in the case of charged black strings \cite{GL94}, among others.     

Recently, an homogeneous black string in AdS was constructed analitically by dressing the extra flat dimensions with minimally coupled scalar fields \cite{ACJO18}. In the context of Lovelock gravity, it was shown that it is possible to construct homogeneous black strings by considering non-minimally coupled scalar fields \cite{Cisterna:2018jsx, Cisterna:2018mww} and also if we  consider that the extra dimensions are dressed with non-minimally coupled p-form fields as seen in  \cite{Cisterna:2020kde, Canfora:2021ttl, Cisterna:2021ckn}, always leading to second order field equations.  It was shown that the solutions in \cite{ACJO18} are stable at linear level under generic perturbations \cite{ACJOCH19}, in contrast of the case of homogeneous black strings in flat space. 

We review first the homogeneous black strings and black $p$-branes solutions of general relativity in $D=d+p$ dimensions with cosmological constant and $p$ massless scalar fields minimally coupled \cite{ACJO18}. For these, the field equations
\begin{eqnarray}
G_{AB}+\Lambda g_{AB}&=&\frac{1}{2}\kappa \sum_{i=1}^{p}\left(\partial_{A}\psi_{i}\partial_{B}\psi_{i}-\frac{1}{2}g_{AB}\left( \partial \psi_{i} \right)^2 \right) , \label{E-K-G-SYS-1} \\
\square \psi_{i}&=& 0 \ , \label{E-K-G-SYS-2}
\end{eqnarray}
where $\kappa=16 \pi G=1$, admit the following $D=d+p$ dimensional  homogeneous black string solution
\begin{eqnarray}
ds^2=-f(r)dt^2+\frac{dr^2}{f(r)}+r^{2}d\sigma_{d-2,\gamma}^2 +\delta_{ij}dy^{i}dy^{j} \label{B-S-sol-1}
\end{eqnarray}
where  $d\sigma_{d-2}^2$ correspond to the line element of a $(d-2)$-dimensional space of constant curvature $\gamma$ and 
\begin{eqnarray}
f(r)=\gamma -\frac{2m}{r^{d-3}}-\frac{2 \Lambda r^2}{\left(d-1 \right)\left(d+p-2 \right) }  \ . \label{lapse-function}
\end{eqnarray}
The solution for the scalar field is given by
\begin{equation}
\psi^{(i)}=\lambda x^{i}  \ ,  \, \, \, \, \ \lambda^2=-\frac{4\Lambda}{d+p-2} \ . \label{scalar-field-sol}
\end{equation}
where $x^{i}$ are Cartesian coordinates along the extended direction.  From these expressions we see that $\Lambda$ has to be negative.

In the next section we will review the linear stability of this solution. 

\section{Stability results}

In \cite{ACJOCH19} it was shown that the black strings constructed in \eqref{B-S-sol-1} are stable at linear level under general perturbations of the spacetime and of the scalar fields. Let us recall this result. For the sake of concretness we consider the five-dimensional case, but the calculations can be carried out in arbitrary dimension. Then, we consider the five dimensional AdS black string solution given by \eqref{B-S-sol-1}, fixing $\kappa=2$ and $\Lambda=-3/2$ such that $\lambda=1$.

Following the originals work \cite{GL93}, we consider a generic, spherically symmetric perturbation for the metric with $g_{AB}\rightarrow g_{AB}+\epsilon h_{AB}$ and
\[h_{AB}=e^{ikz}e^{\Omega t}\begin{bmatrix}
h_{tt}(r) & h_{tr}(r) &0 &0 &h_{tz}(r)  \\
h_{tr}(r) & h_{rr}(r) &0 &0 &h_{rz}(r)  \\
0 &0 &h(r) & 0 & 0 \\
0 &0 &0 &h(r) \sin^2 \theta & 0 \\
h_{tz}(r)& h_{tr}(r)& 0 & 0 & h_{zz}(r) 
\end{bmatrix} \ . \] 
The perturbation for the scalar field is given by 
\begin{equation}
\psi \rightarrow \psi +\epsilon \chi= \psi +\Phi\left(r \right)e^{\Omega t+ik z} \ . \label{sc-pert}
\end{equation}
With the perturbations presented above, we obtain that the linearized Einstein-Klein-Gordon system leads to
\begin{eqnarray}
-\frac{1}{2}\nabla_{L}h_{AB}+\frac{1}{2}g_{AB}h^{CD}R_{CD}+\frac{1}{4}g_{AB}g^{CD}\nabla_{L}h_{CD}+h_{AB}\left(1-\frac{1}{2}R \right)&=&T_{AB}^{(1)} \label{6}  \\
\square \chi -\nabla_{A}h^{AB}-h^{AB}\nabla_{B}\psi -h^{AB}\nabla_{A}\nabla_{B}\psi +\frac{1}{2}\nabla_{A}h\nabla^{A}\psi &=&0 \ , \label{7}
\end{eqnarray}
where $\Delta_{L}$ corresponds to the Lichnerowicz operator of the background spacetime and
\begin{equation}\label{tmunu}
\begin{split}
T_{AB}^{(1)}&=\frac{1}{2}\left(\partial_{A}\psi \partial\chi+\partial_{A}\chi \partial_{B}\psi \right)-\frac{1}{4}g_{AB}g^{CD}\left(\partial_{C}\psi \partial_{D}\chi +\partial_{C}\chi \partial_{D}\psi \right) \\
&-\frac{1}{4}\partial_{C}\psi \partial_{D}\psi\left(h_{AB}g^{CD}-g_{AB}h^{CD} \right) \ .
\end{split}
\end{equation}
Since we are in the presence of matter, it is consistent to impose the following gauge condition $\nabla_{A}h^{AB}=\frac{1}{2}\nabla^{B}h^{C}_{\, \, \, C}$. For this configuration, we obtain that the linearized equation for the scalar field perturbation \eqref{7} takes the form
\begin{equation}
\square \chi -h^{ab}\nabla_{a}\nabla_{b}\psi=0 \ .  \label{8}
\end{equation}
Introducing the separation for the scalar field fluctuation \eqref{sc-pert} we obtain 
\begin{equation} \label{eqhr}
\begin{split}
&\left(r-r_{+}  \right)^2\left(r^2+r_{+}r+r_{+}^2+3 \right)r\frac{d^2 \Phi}{dr^2} -3r^2\left(\left(r-r_{+} \right)k^2+ 3 r\Omega^2 \right)\Phi \\
&+\left(r-r_{+} \right)\left(4r^3-r_{+}^3+6r-3r_{+} \right)\left(r^2+r_{+}r+r_{+}^2 +3\right)\frac{d\Phi}{dr} =0  \ . 
\end{split}
\end{equation}
From this equation, we see that the regular asymptotic behavior at the horizon and at infinity are given by
\begin{eqnarray}
\Phi(r) &\sim& (r-r_{+})^{\frac{r_{+}\Omega}{r_{+}^2+1}}(1+ \mathcal{O}(r-r_{+})) \, \, \, \, \, \textbf{as} \, \, \, \, \, r \rightarrow r_{+}  \ , \\
\Phi(r) &\sim& r^{-\frac{3}{2}-\frac{\sqrt{12 k^2+9}}{2}}(1+ \mathcal{O}(r^{-1})) \, \, \, \, \, \textbf{as} \, \, \, \, \, r \rightarrow \infty \ . 
\end{eqnarray}

In the search for instability, i.e. in the search for positive values of $\Omega$, we see that it is impossible to connect the asymptotic behaviors near horizon and at infinity, therefore, there are no modes with exponential growth in time, and therefore no  GL instability is triggered by the generic perturbations that we have considered. The argument relies on the analysis of a simple second order ODE (see \cite{ACJOCH19}).
  
\section{Non-generic instability}
The equations for the perturbation are still consistent if one fixes $\chi \left(r \right)=0$, i.e if we perturb only the metric. For the scalar mode of the gravitational perturbation $g_{ab}\rightarrow g_{ab}+\epsilon h_{ab}$, where
\[h_{ab}=e^{imz}e^{\Omega t}\begin{bmatrix}
h_{tt}(r) & h_{tr}(r) &0 &0 & 0 \\
h_{tr}(r) & h_{rr}(r) &0 &0 & 0  \\
0 &0 &h(r) & 0 & 0 \\
0 &0 &0 &h(r) \sin^2 \theta & 0 \\
0 &0 &0 &0 & 0 
\end{bmatrix} \ , \] 
we obtain that the $h_{tr}$ satisfies the following equation
\begin{equation}
A\left(r\right)\frac{d^2 h_{tr}(r)}{dr^2}+B\left(r\right)\frac{d h_{tr}(r)}{dr}+C\left(r\right) h_{tr}(r)=0 \ . 
\end{equation} 
with
\begin{equation}\label{A}
\begin{split}
A\left(r \right)&=-\frac{1}{3}\left(\left(k^2+\frac{2}{3}\right) r^{6}+ \left(3\Omega^{2}+3 k^{2}+3\right)r^{4}-r_{+}\left(k^{2}+\frac{4}{3}\right) \left(r_{+}^{2}+3\right)r^{3}-\frac{r_{+}^{2}}{12}\left(r_{+}^{2}+3\right)^2\right)\\
& \times  \left(r^{2}+r r_{+}+r_{+}^{2}+3\right)^{2}\left(-r_{+}+r\right)^{2}r^{2}
\end{split}
\end{equation}
\begin{equation}\label{B}
\begin{split}
B\left(r\right)&= -2 \left(r^{2}+r r_{+} +r_{+}^{2}+3\right)\left(-r_{+}+r \right) r \left(\left( k^{2}+\frac{2}{3}\right) r^{9}+ \left(4\Omega^{2}+4k^{2}+4\right)r^{7} \right. \\
& \quad \left. - \left(k^{2}+\frac{5}{3}\right)r_{+}\left( r_{+}^{2}+3\right) r^{6} +\left( 3\Omega^{2}+3 k^{2}+3\right) r^{5} -\left( k^{2}-\frac{\Omega^{2}}{2} +\frac{3}{2}\right) r_{+}\left(r_{+}^{2}+3\right)r^{4} \right.  \\
& \quad \left. - \frac{r_{+}^{2}}{6} \left(r_{+}^{2}+3\right)^{2}r^{3}-\frac{r_{+}^{2}}{4} \left(r_{+}^{2}+3\right)^{2}r + \frac{r_{+}^{3}}{24} \left(r_{+}^{2}+3 \right)^{3}\right) 
\end{split}
\end{equation}
and
\begin{equation}\label{C}
%\begin{eqnarray*}
\begin{split}
C\left(r \right)&=\left(-\frac{8}{9}+k^{4}-\frac{2k^{2}}{3}\right)r^{12}+\left(6 k^{4}+\left(6\Omega^{2}-3\right) k^{2}-6\Omega^{2}-8\right)r^{10}+\frac{r_{+}^{4}}{36}\left(r_{+}^{2}+3\right)^{4}\\
&-2r_{+}\left(k^{2}-\frac{11}{6}\right)\left(k^{2}+\frac{4}{3}\right)\left(r_{+}^{2}+3\right)r^{9}+\left(9\left(k^{2}+\Omega^{2}-\frac{2}{3}\right)\right)\left(\Omega^{2}+k^{2}+1\right)r^{8}\\
&-6r_{+}\left(k^{4}+\left(\Omega^{2}+\frac{1}{2}\right)k^{2}+\Omega^{2}-\frac{4}{3}\right)\left(r_{+}^{2}+3\right)r^{7}+\left(18 k^{2}+18\Omega^{2}+18\right)r^{6}\\
&+\left(\left(-\frac{1}{3}+k^{4}+\frac{3}{4}k^{2}\right)r_{+}^{6}+\left(-2+6k^{4}+\frac{9}{2}k^{2}\right)r_{+}^{4}+\left(-3+9 k^{4}+\frac{27}{4}k^{2}\right)r_{+}^{2}\right)r^{6}+\\
&-21 r_{+}\left(r_{+}^{2}+3\right)\left(k^{2}+\frac{4}{7} \Omega^{2}+\frac{6}{7}\right)r^{5}+\frac{33}{4}r_{+}^{2}\left(k^{2}+\frac{1}{11}\Omega^{2}+\frac{38}{33}\right)\left(r_{+}^{2}+3 \right)^{2}r^{4}\\
&-\frac{13}{12}r_{+}^{3}\left(k^{2}+\frac{4}{3}\right)\left(r_{+}^{2}+3\right)^{3}r^{3}+\frac{3}{2}r_{+}^{2}\left(r_{+}^{2}+3\right)^{2}r^{2}-\frac{1}{2}r_{+}^{3}\left(r_{+}^{2}+3\right)^{3}r
\end{split}
%\end{eqnarray*}
\end{equation}
The asymptotic behavior of $h_{tr}$ near the horizon (when $r\rightarrow r_{+}$ ) is given by
\begin{equation}
h_{tr}\left(r \right)=c_{1}\left(-r_{+}+r \right)^{-1
+\frac{\Omega r_{+}}{r_{+}^2+1}}\left(1+\mathcal{O}\left(r-r_{+} \right) \right)+c_{2}\left(-r_{+}+r \right)^{-1-\frac{\Omega r_{+}}{r_{+}^2+1}}\left(1+\mathcal{O}\left(r-r_{+} \right) \right)
\end{equation}

For unstable modes $\Omega > 0$, therefore the second branch diverges at the horizon. Note that for $\Omega < \frac{1+r_{+}^2}{r_{+}}$ even the first branch diverges at the horizon. As in the original GL case, we will actually find unstable modes in this regime. It turns out that the divergence of these modes near the horizon, is an artifact of the coordinates, and one can show that in Kruskal-Szekeres coordinantes such modes are regular. 
  
At the other boundary, when $r \rightarrow \infty$, the asymptotic behavior is given by
\begin{equation}
h_{tr}\left(r \right)=\tilde{c}_{1}\left(\frac{1}{r} \right)^{\frac{5}{2}+\frac{\sqrt{9+12k^2}}{2}}\left(1+\mathcal{O}\left(\frac{1}{r} \right) \right)+\tilde{c}_{2}\left(\frac{1}{r} \right)^{\frac{5}{2}-\frac{\sqrt{9+12k^2}}{2}}\left(1+\mathcal{O}\left(\frac{1}{r} \right) \right)
\end{equation}

It is interesting to notice there is a range of momenta $k$ along the extended direction for which both branches are regular at infinity. Following the standard approach in AdS, we will impose Dirichlet boundary conditions at infinity , i.e we will require $\tilde{c}_{2}=0$ which kills the slow branch. 

Before implementing the method that will lead to the spectrum, it is useful to introduce the following change of variables
\begin{equation}
h_{tr}\left(r \right)=\left(r-r_{+}\right)^{-1+\frac{r_{+}\Omega}{r_{+}^2+1}}r^{1-\frac{r_{+}\Omega}{r_{+}^2+1}}r^{-\frac{5}{2}+\frac{\sqrt{9+12k^2}}{2}}\tilde{h}\left(r \right) \ .
\end{equation}
With this redefinition, $h(r)$ inherits the following asymptotic behavior at the horizon
\begin{equation}
\tilde{h}\left(r \right) \sim c_{1}\left(1+\mathcal{O}\left(r-r_{+}\right) \right)+c_{2}\left(r-r_{+} \right)^{-\frac{2r_{+}\Omega}{r_{+}^2+1}}\left(1+\mathcal{O}\left(r-r_{+}\right) \right) \ ,
\end{equation}
and at infinity $r\rightarrow \infty$
\begin{equation}
\tilde{h}\left(r \right) \sim \tilde{c}_{1}r^{-\sqrt{9+12k^2}}\left(1+\mathcal{O}\left(1/r \right) \right)+\tilde{c}_{2} \ .
\end{equation}
and therefore we need to set $c_{2}=0$ and $\tilde{c}_{2}=0$ at the horizon and at infinity, respectively.

To obtain the spectrum, it is convenient to introduce the coordinate $p=\frac{r-r_{+}}{r}$, which maps $r \in ]r_{+}, +\infty[$ to $p \in ]0,1[$. By imposing the appropriate boundary conditions, and using a power series solution for the perturbation, we found positive values of $\Omega$, indicating an exponential growth of the perturbation over time. For $r_{+}=0.1$, we obtain the spectrum of instabilities shown in fig.1 and fig.2 depicts the behaviour of the perturbation at several orders in the serie.

%\begin{verbatim}
\begin{figure}[h!]
\begin{center}
%\begin{small}
\includegraphics[scale=0.70]{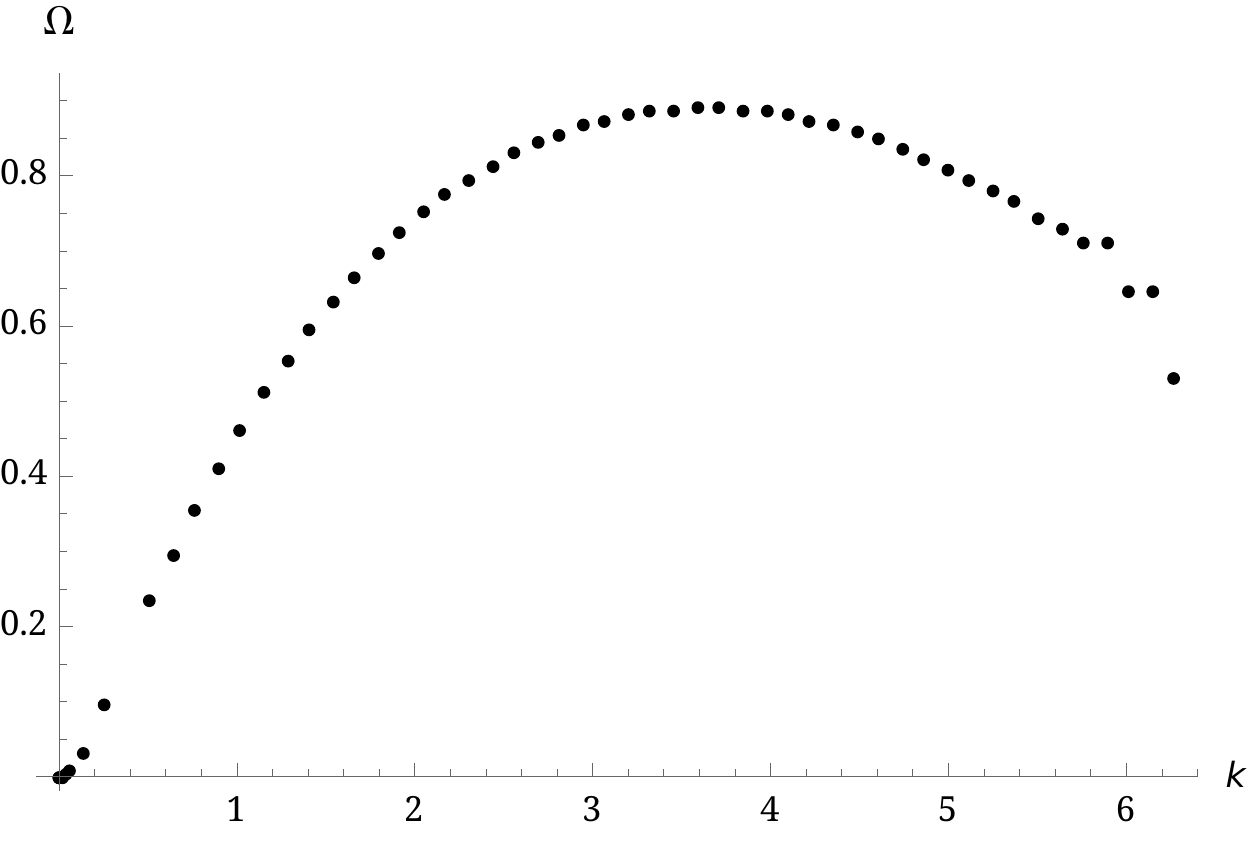} \qquad
%\end{small}
\end{center}
\caption{\label{label}The spectrum of the instability for $r_{+}=0.1$. This was obatined with 30 orders in the power series. For higher values of the wave number $k$, the convergence of our numerical scheme breaks down. }
\end{figure}

\begin{figure}
\begin{center}
\begin{small}
\includegraphics[scale=0.70]{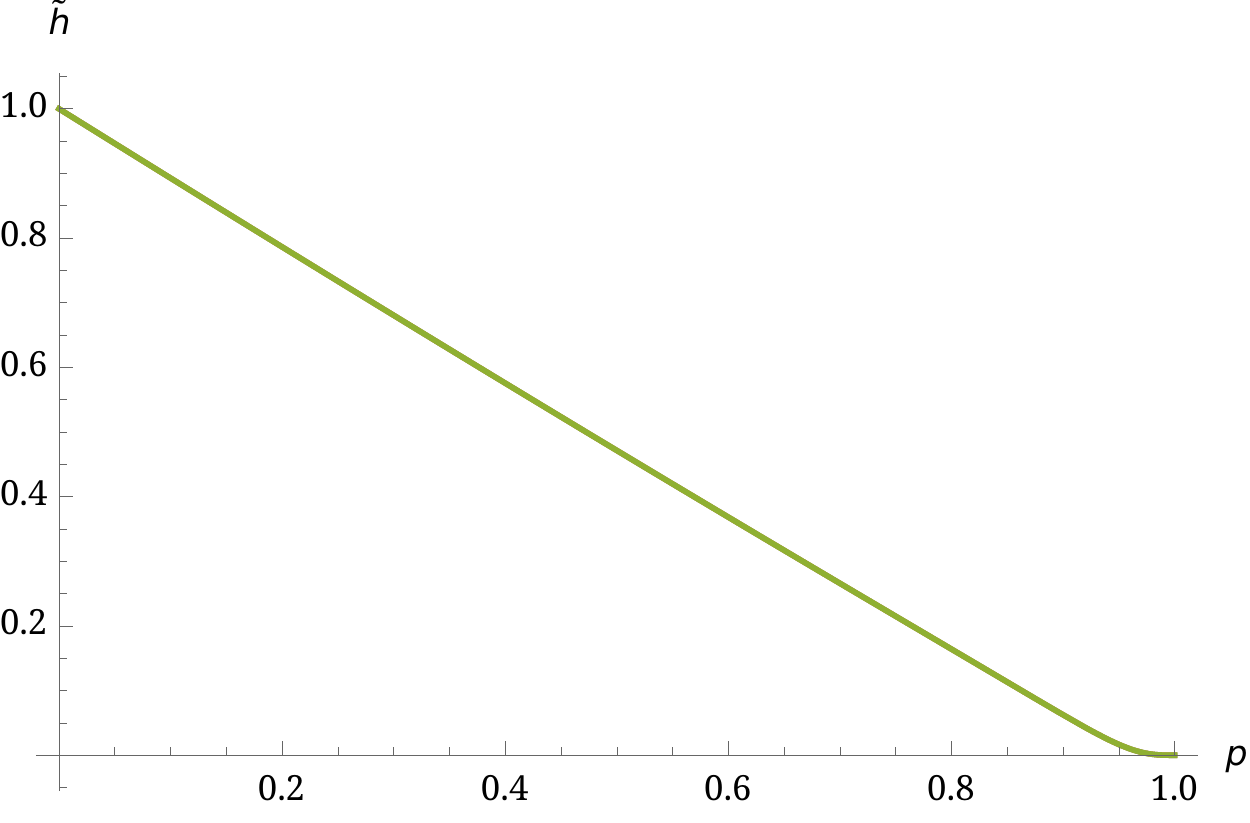}
\end{small}
\end{center}
\caption{\label{label}A plot of the behavior of the metric perturbation $\tilde{h}$ as a function of the radial coordinate $p$}
\end{figure}
%%\end{verbatim}

\section{Further remark}
Here we have extended the results about the stability of an homogeneous black string in AdS. We have shown that a non-generic perturbation of the homogeneous black string solution \eqref{B-S-sol-1} leads to a GL instability. It is important to mention that although for the case where the scalar field perturbation vanishes, the GL instability is established. We must remark that this is a consistent, non-generic perturbation.

In this new version it is important to mention that in \cite{Dhumuntarao:2021gdb} an extended analysis was presented, which was developed in parallel. We thank the authors for contacting us and sharing their preprint.

Finally let us mention that a recent large D analysis performed in \cite{Li:2021lkw} supports the stability results for generic perturbations at the non-linear level. 

\section*{Acknowledgments}
%\verb"\ack"
This work extends some previous results obtained in collaboration with A. Cisterna and J. Oliva, whom I would like to thank for the numerous and enlightening debates. This work was funded by the National Agency for Research and Development (ANID) /  Scholarship Program / BECA DE DOCTORADO NACIONAL/2017 - 21171394 and it has been partially fundend by Fondecyt Grant 1181047.

\end{document}